# 25 YEARS OF COSMIC MICROWAVE BACKGROUND RESEARCH AT INPE

**Carlos Alexandre Wuensche and Thyrso Villela**

*Divisão de Astrofísica - Instituto de Pesquisas Espaciais - INPE*
*Av. dos Astronautas,1758 – 12201-970, São José dos Campos-SP, Brasil*

## ABSTRACT

This article is a report of 25 years of Cosmic Microwave Background activities at INPE. Starting from balloon flights to measure the dipole anisotropy caused by the Earth's motion inside the CMB radiation field, whose radiometer was a prototype of the DMR radiometer on board COBE satellite, member of the group cross the 90s working both on CMB anisotropy and foreground measurements. In the 2000s, there was a shift to polarization measurements and to data analysis, mostly focusing on map cleaning, non-gaussianity studies and foreground characterization.

## INTRODUCTION

The cosmic microwave background radiation (CMB) is one of the most important cosmological observables presently available to cosmologists. Its properties can unveil information, among others, about the inflationary period, the overall composition of the Universe ($\Omega_0$), the existence of gravitational waves, the age of the Universe and other parameters related to the recombination and decoupling era (Hu and Dodelson, 2002). These observables are critical to understand the physical processes accounting for the formation and evolution of the Universe. The CMB is observed from a few GHz to a few hundreds of GHz. It also observed in various angular scales, varying from less than 1 arcmin to many degrees, each range of scales encoding information about specific physical processes from the early (or not so early) Universe. Its properties (intensity, anisotropies and polarization) can be studied from the ground, on board stratospheric balloons and satellites and are strongly hampered by the so-called foreground contaminants, constituted by radio point sources and Galactic emission.

It is expected that the next space missions to measure the anisotropy and the polarization of CMB will mostly deal with foreground systematics, since the sensitivity obtained from years in space can only increase by a $(t)^{-1/2}$ factor and the technology used for detectors are quickly approaching the quantum limit efficiency (see, e.g., CMB Task Force 2006, hereafter CMBTF2006). Also, most detectors used in the present generation of instruments and projected for the next one are in the 10 - 1 microK.sqrt(Hz) sensitivity (ref). On the other hand, foreground emission is arguably the last frontier to CMB studies, since the details of synchrotron, free-free and dust emission, as well as their spectral and spatial distributions are still poorly understood. Presently, the foreground emission from the abovementioned processes accounted for in many different CMB missions (see, e.g., CMB F2006) between 30 and 600 GHz can contribute at the same intensity level or above, than the E and B polarization modes of CMB.

INPE's Cosmology Group (hereafter CG) has contributed to many aspects of CMB studies in the last 25 years and is presently devoted to foreground emission and total intensity measurements in the experimental side (see, e.g., Tello et al., 2007; Levy et al., 2008; Kogut et al., 2009; Fixsen et al., 2009; Seiffert et al., 2009). On the phenomenological and data analysis aspects, the group has contributed in the CMB anomalies and asymmetries aspects, as well as in the search of global isotropy of the Universe (see, e.g., Abramo et al. 2006; Abramo, Sodré and Wuensche, 2006; Bernui and Villela, 2006; Bernui et al., 2007; Bernui, Tsallis and Villela, 2007; Bernui, Ferreira and Wuensche, 2008; Bernui 2008).

This paper will describe briefly the contributions of the CG to CMB studies, including only the most significant results from the 80s, 90s and 2000s. Section 2 presents the CMB history and INPE's CG participation. Section 3 includes our contribution to foreground studies. Section 4 contains our data analysis and phenomenology contributions and the technological contributions are in Sec. 5. We summarize this article in Section 6.

## INPE AND CMB HISTORY

In the early 80's, three issues were attacked by CMB researchers: the expected blackbody characteristics, of CMB, the expected dipole signal from our movement though CMB photons and the temperature fluctuations left on CMB as a result of the primordial density fluctuation,s due to the coupling between photon and matter before the recombination era, at z ~ 1400 - 1100.

### The era pre-COBE

A series of balloon flights were performed in the USA and in Brazil to nail down the dipole characteristics, which was not a cosmological signal, and to search for the cosmological CMB quadrupole signal. The combined result from these two flights were reported by Lubin and Villela (1986) and Lubin et al (1985), with a dipole intensity of 3.44 ± 0.17 mK and a direction of RA = 11.2 h, Dec = -6°. The reported quadrupole upper limit was $7 \times 10^{-5}$ microK. The instrument used in these flights was a prototype of one of the DMR (Differential Microwave Radiometer) used onboard COBE satellite, which was launched in 1989.

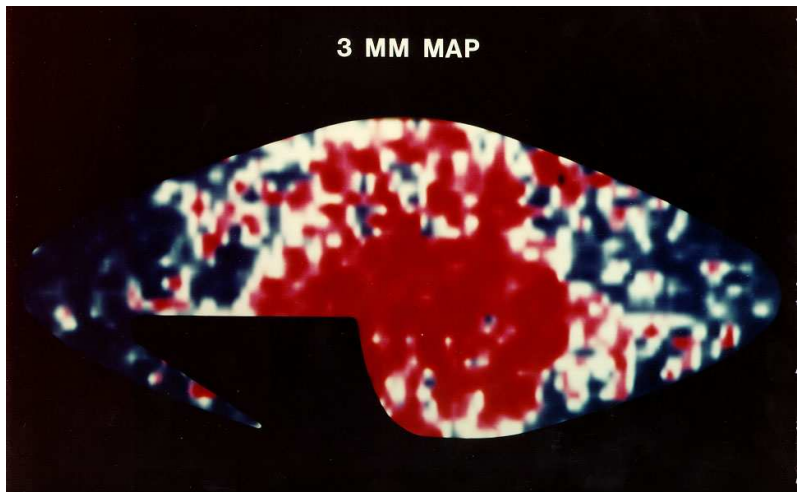

Fig. 1 - Dipole anisotropy map (in celestial coordinates) covering 85% of the sky. The black region in the lower left part of the map was not covered during the two observation campaigns.

**The COBE times**

COBE was launched in 1989 and was the first NASA satellite fully dedicated to measure the CMB. It carried three experiments onboard: the FIRAS (Far InfraRed Absolute Spectrophotometer) to measure the CMB spectrum, operating from 60 to 600 GHz, The DMRs (Differential Microwave Radiometers) were designed to measure the angular distribution and operated in 31, 53 and 90 GHz. DIRBE (Diffuse InfraRed Background Explorer) was designed to measure mostly the dust emission in the far infrared band of the eletromagnetic spectrum (from 240 μm to 1.25 μm). The COBE mission ended in 1994 and was very successful, setting milestones for the next generation of balloon and ground experiments during the 90s.

The blackbody shape of CMB temperature spectrum was reported weeks after launch (Mather et al., 1990, Figure 2), showing an almost perfect blackbody shape. The expected large-scale temperature fluctuations were reported a little later by Smoot et al. (1992). The COBE maps, from a full sky map to clean, CMB only map, can be seen in Figure 3. The interpretation of COBE results were discussed in a companion paper by Wright et al. (1992). However, the DMR observed the sky with a large horn (about 7 degrees FWHM) and did not probe intermediate and small angular scales ($\theta < 2°$), where the signature of the density fluctuations in the form of acoustic peaks should be found.

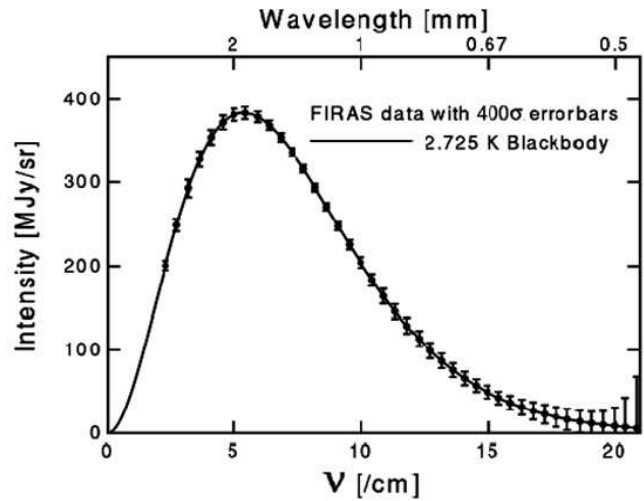

Fig. 2 - The FIRAS measurements on top a theoretical blackbody curve. Note the amplification of the error bars in order to be noticed in the graph.

One of the COBE science team members, Philip Lubin, then at University of California Santa Barbara, proposed a series of experiments in the late 80s, to measure these medium scales. Together with Paul Richards, at the University of California, Berkeley, they started a series of experiments to search for CMB anisotropy in medium angular scales. The result were 4 balloon flights with bolometers as their detectors (the ACME-MAX series: Alsop et al., 1992; Gundersen et al., 1993; Meinhold et al., 1993; Clapp et al., 1994; Devlin et al., 1994; Tanaka et al., 1996; Lim et al., 1996), and 4 campaigns at the South Pole, using HEMT detectors (the ACME-SP series: Meinhold and Lubin, 1990; Gaier et al., 1992; Schuster et al., 1993; Gundersen et al., 1995). Some of their results, detecting medium-scale temperature fluctuations, can be seen in Figures 4, 5 and 6.

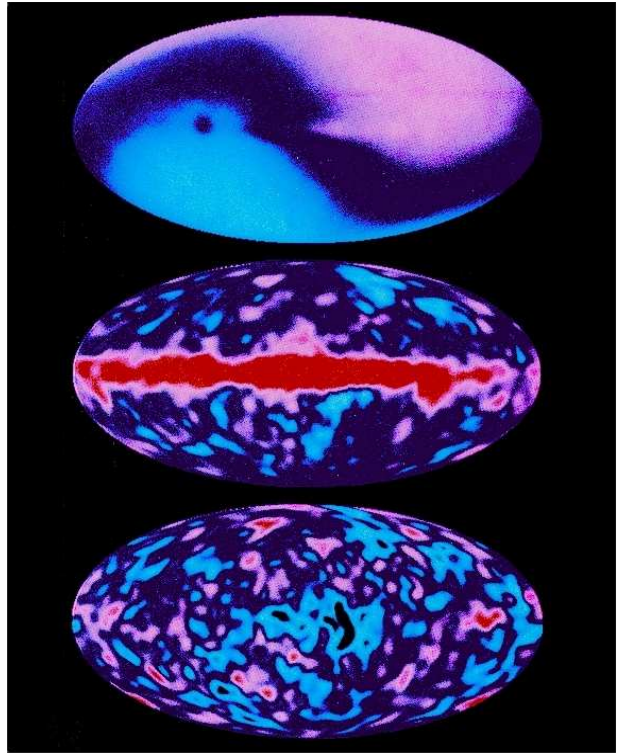

Fig. 3 - COBE anisotropy maps. The top map contains the dipole and the Galaxy emission on top of the CMB anistropies; the center map had the dipole removed, containing CMB and Galaxy; the bottom map contains CMB fluctuations only.

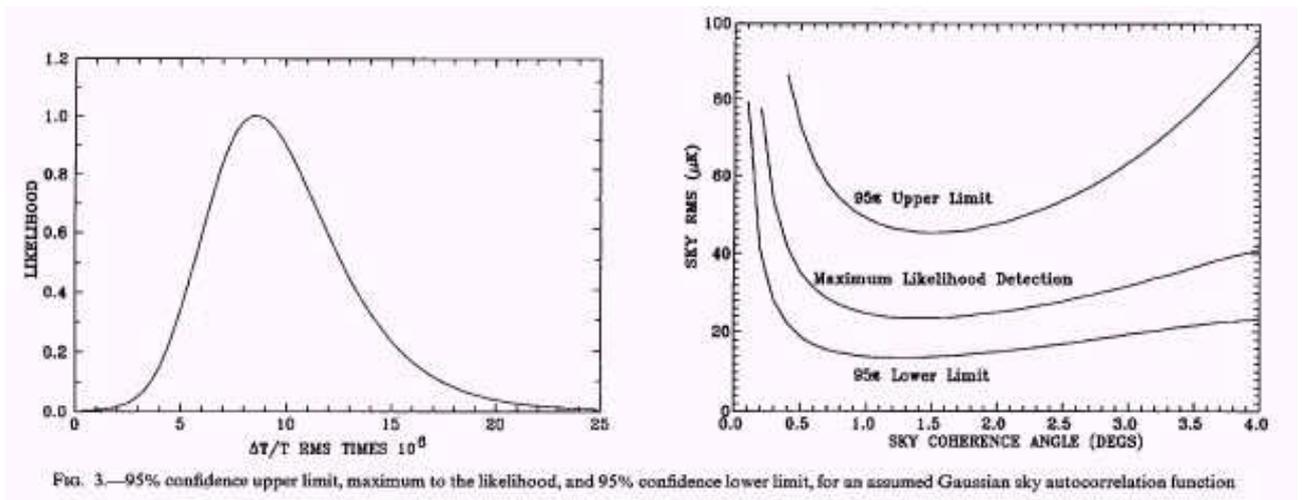

FIG. 3.—95% confidence upper limit, maximum to the likelihood, and 95% confidence lower limit, for an assumed Gaussian sky autocorrelation function

Fig. 4 - Results from Schuster et al (1993). Note the picture in the right, with the autocorrelation function results used to verify the results and claim the detection as done in this work.

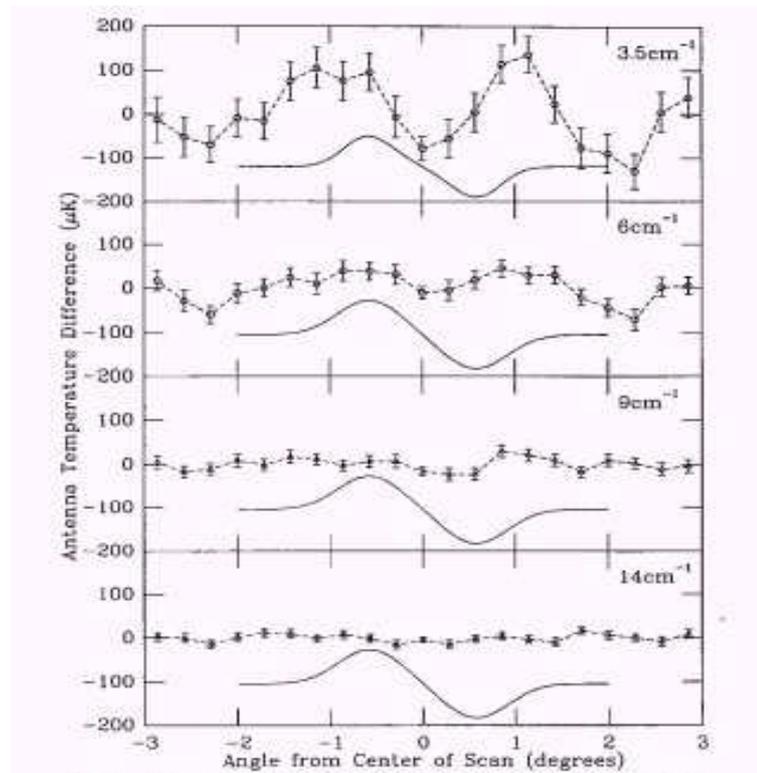

Fig. 5 - Results from Devlin et al. The CMB excess claimed as a detection can be noticed in the upper part of the figure.

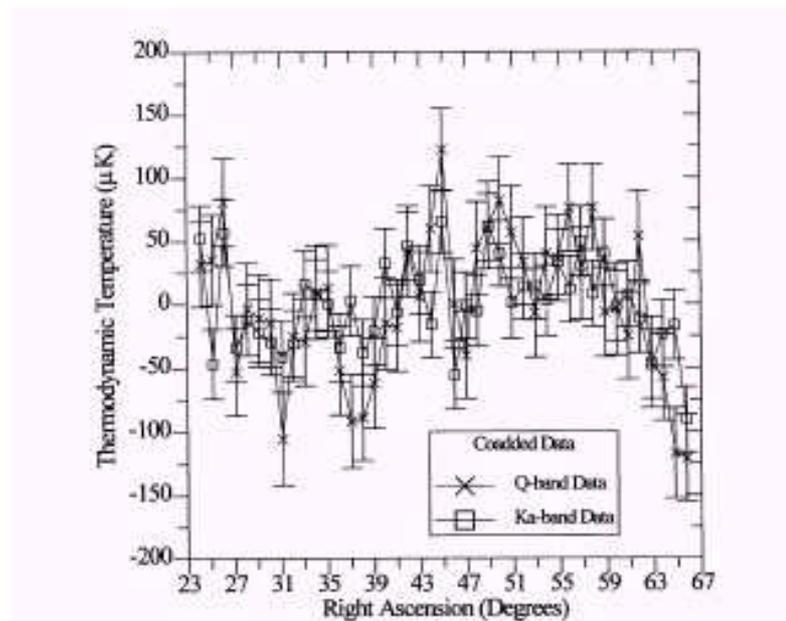

Fig. 6 - The ACME-SP 1995 from Gundersen et al. These data points were one of the first to be used to delimit the acoustic CMB peak.

In the late 90s, the UCSB group developed a new telescope concept, using a large (2.6 m diameter), flat rotating mirror as a tool to quickly cover large regions of the sky, and flew this instrument on a balloon. Since it carried HEMT detectors but flying onboard a stratospheric balloon, it was named HACME, standing for Hemts on ACME (Staren et al. 2000; Tegmark et al., 2000, Figure 7). HACME made one of the first maps of CMB anisotropy in medium angular scales and was a prototype for another new telescope: BEAST (Background Emission Anisotropy Scanning Telescope), to be described in the next section.

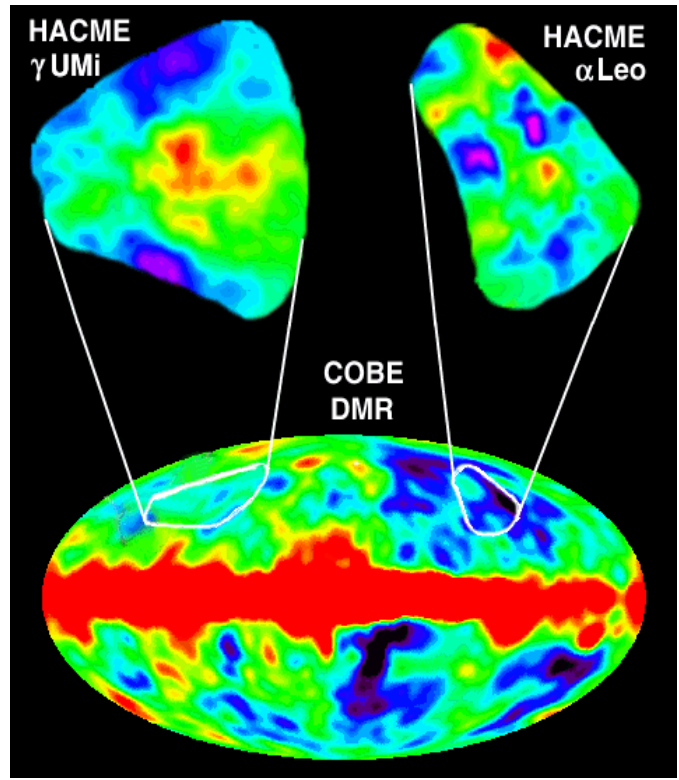

Fig. 7 - A high resolution CMB map, made by the HACME experiment, compared to the COBE all-sky map.

**The 2000s**

The BEAST telescope operated on the ground, in the White Mountain Research Station (USA) and incorporated two carbon fiber large mirrors (paraboloid of 2.2 m and ellipsoid of 0.8 m in the semi-major axes) on board a gregorian mounting with an off-axis optic (Figure 8). It also featured a new optical design to minimize sidelobe contributions (Figueiredo et al., 2005; Childers et al., 2005), building on the experience of its predecessors. BEAST measured the CMB anisotropies and the Galaxy contribution around the north celestial pole (Meinhold et al. 2005; Mejia et al. 2005) and was the first instrument of the group to measure the CMB power spectrum and put limits in the cosmological parameters (O'Dwyer et al., 2005). Figure 9 shows the area covered by BEAST (about 4% of the sky) compared to a full sky WMAP map.

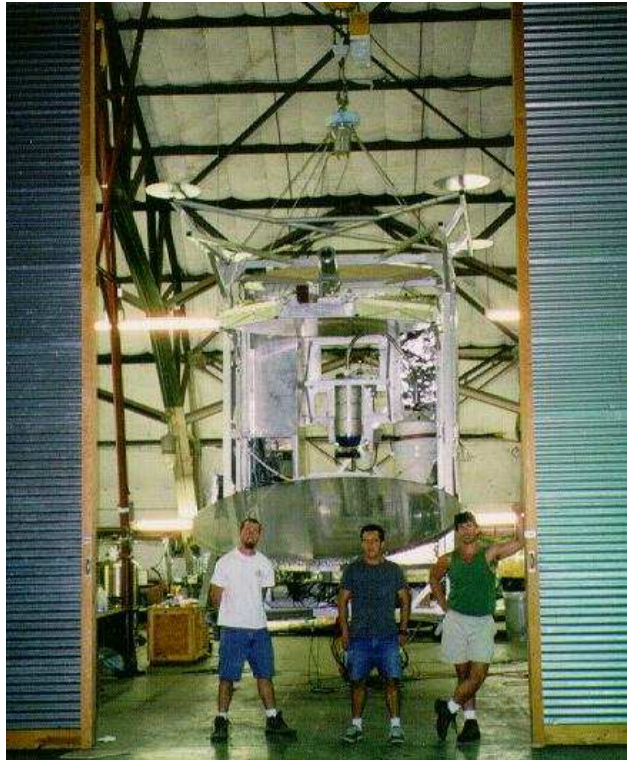

Fig. 8 - BEAST in its dome in White Mountain. The size of the instrument and components can be figured out by comparing with investigators in front of the gondola. They are about 1,80 m tall.

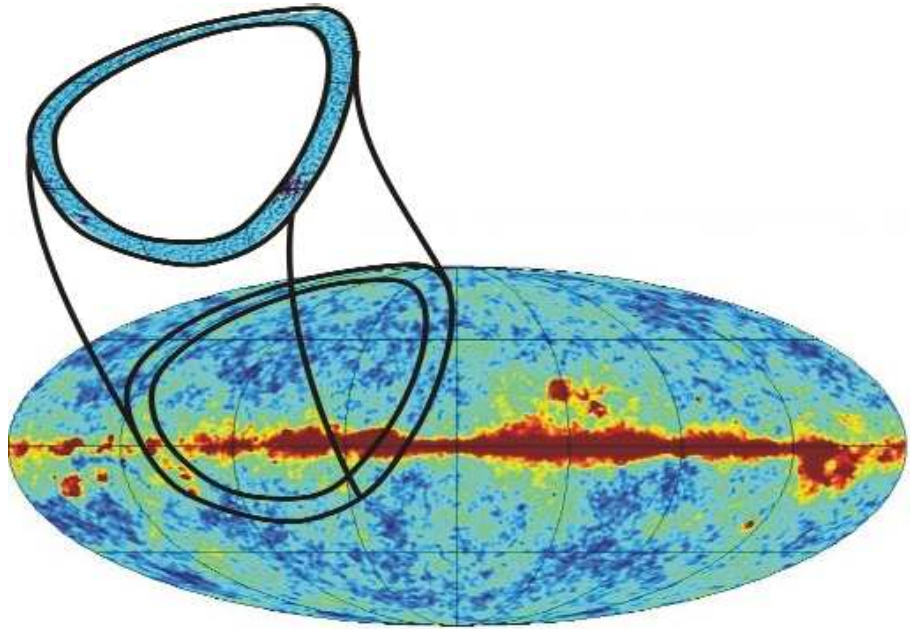

Fig. 9 - Superposition of BEAST on top a combined WMAP map including the Galaxy.

The natural extension of BEAST would be an instrument with polarization measurement capabilities and this came with the WMPol (White Mountain Polarimeter), also operating at White Mountain Research Station. The results were not as good as the previous one, due to instrumental and sensitivity limitations. The instrument took data at 42 GHz and reported upper limits for CMB E-mode polarization of 14 microK (Levy et al. 2008). It was intended as a prototype for a more sofisticated "foreground-cleaner" instrument, setup to measure CMB foregrounds between 10 and 20 GHz, and also to be sensitive to B-mode polarization. This experiment, called COFE (COsmic Foreground Explorer), is described by Leonardi et al. (2006).

**FOREGROUND STUDIES**

INPE's CG has been involved in foreground studies since the early 90's, when the GEM (Galactic Emission Mapping) project started. Its goal: to map synchrotron radiation at 408 MHz, 1,465 MHz, 2.3 GHz, 5 GHz and 10 GHz along 60-deg wide declination bands from several locations with a single dish. Using the 408 MHz measurements as cross-check with the Haslam survey (Haslam et al 1982) and the novel observational strategy of a rotating dish, the GEM experiment began its search for the spatial and spectral distribution of foreground templates without the baseline inhomegeneities that plagued combinations of earlier low frequency surveys. Another checks could be done against the 1420 MHz Reich and Reich survey (1986) and the Jonas 2.3 Gz survey (2003), In addition, Stokes U and Q polarization components have been implemented for the 5 and 10 GHz experiments, which avoid Faraday rotation to a cosmological level of interest for CMB studies.

GEM started its operations at 408 and 1,465 MHz in Bishop, California (USA) in 1993 and it spent a brief, but intense, campaign in Villa de Leyva, Colombia, (Torres et al 1996) where 2.3 GHz observations were added. Mapping of the Southern Sky began after its arrival in Brazil in 1998 (Tello et al 2000, 2007) and recently polarization measurements at 5 GHz became a very welcome reality (Ferreira 2008).

INPE's CG participation in the development of the GEM project has been crucial since its early stages, from the construction of receivers at 1,465 MHz (Tello 1997) and 5 GHz (Ferreira 2008) to a more efficient design of the ground screen and shield (Tello et al 1999, 2000) and the adaptation of a cryogenic system for the 5 and 10 GHz receivers. Understanding ground contamination has been one of GEM's main assets in producing a reliable baseline for maps of the radio continuum at low ferquencies (Tello et al 1999). Polarization measurements will also benefit of newly implemented control and cryogenic systems.

In 2009 the CG will completely refurbish the optical design of the GEM dish to accomodate the challenging requirements of polarization measurements at 10 GHz, as well as the design and fabrication of the 10 GHz receiver, which will be completely fabricated in Brazil.

**CMB DATA ANALYSIS**

Additionally, the group has devoted some time to do some phenomenology and worked in CMB data analysis as well, both to produce science results and to propose new analysis tools. In the early 2000's there were some efforts to search for non-gaussianity in the CMB anisotropies, addressed in a series of papers by Andrade, Wuensche and Ribeiro (2004, 2005, 2006). This deviation from gaussianity can be tested with a non-extensive statistics approach proposed by Tsallis (1988). Bernui, Tsallis and Villela (2006, 2007) did such test and reported a very small degree of non-extensivity in the CMB data using 1- and 3-year WMAP data.

After the WMAP results were released (Bennett et al. 2003, Hinshaw et al. 2007), the questions about the isotropy and asymmetry of CMB angular distribution was studied using a statistical tool, the PASH, proposed by Bernui and Villela (2006). These papers examined both the isotropy and the symmetries of CMB anisotropies (Bernui, Ferreira and Villela 2004; Bernui 2005; Bernui et al 2006; Abramo et al. 2006, Abramo, Sodré and Wuensche 2006).

Using the same idea, Bernui, Ferreira and Wuensche (2008) explored the isotropy of the GRB distribution in the Universe, showing no major evidence of anisotropy in the data.

## TECHNOLOGICAL DEVELOPMENTS

INPE's CG has taken part in a large number of CMB missions, contributing, in many of them, towards the technological side of the instrument, besides data analysis and science discussions. Table 1 shows some of CG contributions since 1990, and in which stage there was a major involvement.

Table 1 - CG contributions since 1990.

| Part | Study | Design | Fabr. | Test | Oper. | Year |
|------|-------|--------|-------|------|-------|------|
| BEAST Horns | | | X | X | | 2001 |
| Waveguides and transitions | | X | X | | | 2001 |
| ARCADE horns and transitions | | X | X | X | | 2005 |
| GEM Dewar | X | X | X | X | X | 2006 |
| GEM fences | X | X | X | X | X | 2001 |
| 1.465 and 5 GHz receivers | X | X | X | X | X | |
| Optical project for HACME and BEAST | X | X | X | | | 1996 |
| Microwave simulations for WMPOL and GEM | X | X | X | X | | 2005-2007 |
| 5 GHz OMT | X | X | X | X | | 2006 |

## CONCLUSIONS

Since its creation, in the late 80s, the CG at INPE has been deeply involved in all aspects of CMB measurements and CMB foregrounds, particularly synchrotron emission. There were collaborations covering ground, balloon and satellite missions, mostly with groups from Berkeley and Santa Barbara (USA) and a deep involvement of all the scientists and students from the group.There also efforts devoted to CMB data analysis and phenomenology addressing issues in datasets produced by instruments in which the CG was involved. These efforts attacked mostly non-gaussianity aspects of CMB distribution and its global isotropy and asymmetry.

Summing up all contributions, INPE's CG produced a total of 54 papers in about 25 years of existence, including a major gap, from 1987 to 1997, where most of the PhD degrees produced in the group were concluded. The group produced, as of December 2008, a total of 9 PhD theses and 9 MSc dissertations, some of which now hold permanent positions at universities and research centers in Brazil and abroad, and still keep some level of contact with the science scene at INPE.

Presently the group is mostly involved in the development of a 10 GHz receiver for the GEM instrument and is working on the data analysis of the ARCADE instrument, a collaboration between Goddard Space Flight Center/NASA, Jet Propulsion Lab/NASA, University of California Santa Barbara, University of Maryland, all in USA, and INPE.